\documentclass[aps, pra, showpacs, twocolumn]{revtex4}

\usepackage{graphics}
\usepackage{graphicx}
\usepackage{float}
\usepackage{amssymb}
\usepackage{amsmath}
\usepackage{bm}       
\usepackage{mathrsfs} 
\usepackage{ae}
\usepackage{natbib}

\newcommand{\Partial}[2]{\frac{\partial {#1}}{\partial {#2}}}
\newcommand{\dxi}{\int\!\text{d}x\ }

\newcommand{\dki}{\int\text{d}k\ }
\newcommand{\od}{_\text{1D}}
\begin{document}

\title{Tunnelling couplings in discrete lattices, single particle band
structure and eigenstates of interacting atom pairs}

\author{Rune Piil}
\email{piil@phys.au.dk}

\author{Klaus M\o{}lmer}

\affiliation{Lundbeck Foundation Theoretical Center for Quantum System
  Research, Department of Physics and Astronomy, University of
  Aarhus, DK-8000 Aarhus C, Denmark}

\date{\today}

\pacs{03.75.Lm, 03.65.Ge}

\begin{abstract}
  By adjusting the tunnelling couplings over longer than
  nearest neighbor distances it is possible in discrete lattice models
  to reproduce the properties of the lowest energy band of a real,
  continuous periodic potential. We propose to include such terms in
  problems with interacting particles and we show that they have
  significant consequences for scattering and bound states of atom
  pairs in periodic potentials.
\end{abstract}

\maketitle

\section{Introduction}

Spatially periodic potentials induced by off-resonant laser fields
have been applied in numerous studies to modify the dynamics of
ultra-cold bosonic gases. In the ``mean-field'' regime with many
atoms, the potential changes the transport properties and effective
mass of the atoms \cite{PRL:140406,PhysRevA.64.061603,KarenMarie02},
in a dilute gas the lattice enables tuning of the ratio between
kinetic and interaction energy and observation of a Tonks gas
\cite{Nature:429-277,kinoshita2004ood}, and in the limit of well
confined atoms in deep lattice wells collisionally induced collapses
and revivals of the matter wave mean field has been observed
\cite{Nature:419-51}. The observation of the transition from a
superfluid to a Mott insulator is a hallmark experiment on a zero
temperature phase transition made possible by the use of an optical
lattice \cite{PhysRevB.40.546,PRL:81-3108,Nature:415-39}.

To deal properly with the many-body aspects of the quantum state, it
has been useful to approximate the motion in the periodic potential by
a discrete lattice model, specifying for a single atom the atomic wave
function only on positions corresponding to the potential minima of
the true potential. The kinetic energy gives rise to tunnelling
couplings between the sites, whereas the potential energy is described
by local values of the potential on each site. If the potential
attains the same value on each site, the eigenvectors of the
Hamiltonian are characterized by amplitudes that experience a fixed
phase rotation as one steps through the lattice. These states are
(quasi-)momentum eigenstates, and their energies depend on the
tunnelling coupling coefficients.  The discrete lattice model with
only nearest neighbor tunnelling couplings does not reproduce
important features in the momentum dependence of energy in the lowest
energy band of a real potential. In this paper we show that it is
relatively easy to remedy this problem by incorporation of ``long
distance'' tunnelling couplings, and we propose that such couplings be
incorporated in studies of many-body dynamics in periodic potentials,
since binding and collisional interactions may depend crucially on the
dispersion properties of atoms in the periodic environment.

As a particular example we consider two interacting atoms, and in
particular the phenomenon of repulsively bound atom pairs in optical
lattices \cite{winkler2006rba, denschlag2006eap, PhysRevE.66.016130,
  2006JPhA...39L.667M}. The stability of a repulsively bound pair may
be understood qualitatively from the observation that the repulsive
interaction gives two atoms an energy that lies within the energy band
gap of free motion in the periodic potential, forcing the atoms to
stay together in a discrete bound state lying above the energy
continuum of scattering states. The lack of dissipative channels,
unlike, e.g., the phonons in real crystals, makes the observation of
repulsively bound atom pairs possible in optical lattices
\cite{winkler2006rba}.

The paper is structured as follows. In Sec. II, we analyze the eigenstates of a single particle in a
periodic potential, and we show that the lowest energy band can be obtained also from a discrete
lattice model with suitably selected tunnelling couplings. In Sec. III, we turn to the description of
two atoms with a short range interaction, and we identify the continuum of scattering states in the
lowest Bloch band. In Sec. IV we identify the discrete states following from the attractive and
repulsive binding mechanisms, and we show that these states have a number of physical properties that
depend markedly on our more precise description of the band structure of the system. Sec. V concludes
the paper and discusses possible consequences of incorporating its results in the treatment of
many-body problems, e.g., via the Bose-Hubbard Hamiltonian.

\section{Single-particle dynamics}

We consider an atom with mass $m$ that moves in an off-resonant standing wave laser field with wave
length $\lambda$. Due to the light induced energy shift, the atom experiences a potential
$V(x)=V_0\cos(2\pi x/a)$ with period $a=\lambda/2$ and potential strength $V_0$ proportional to the
light intensity and inversely proportional to the optical detuning from atomic resonance. According to
Bloch's theorem, the Hamiltonian
\begin{equation}
  \label{eq:SingleHamiltonian}
  \mathscr{H}=
  -\frac{\hbar^2}{2m}\Partial{^2}{x^2}+V(x)
\end{equation}
is diagonalized by wave functions $\phi_{nk}$, which are products of plane wave factors $\exp(ikx)$
and periodic functions. The energies of these states $E^n(k)$ describe energy bands as the
quasi-momentum $k$ is varied across the first Brillouin zone $[-\pi/a,\pi/a]$. It is convenient to
represent the energy in units of the lattice recoil energy $E_R=\frac{h^2}{2m\lambda^2}$. $n$ is the
band index, and in the following we shall restrict our analysis to the lowest band with $n=1$ and omit
the quantum number $n$ from our equations. An alternative basis for the quantum states in the lowest
band is the spatially localized Wannier functions
\begin{equation}
  \label{eq:WannierFunction}
  w_{j}(x)=\frac{1}{\sqrt{M}} \sum_{k} e^{ikaj}\phi_{k}(x)
\end{equation}
describing atoms localized in the $j^{th}$ well in the lattice (to
ensure constructive interference at the $j^{th}$ well we need a phase
convention and choose $\phi_{k}(x=0)$ to be real and positive for all
$k$). We have, for convenience, assumed a finite number $M$ of lattice
periods and periodic boundary condition on the entire system.

By evaluating all matrix elements
\begin{equation}
  \label{eq:Constants}
  \begin{split}
    J_d&\equiv -H_{j,j+d}= -\langle w_j|\mathscr{H}|w_{j+d}\rangle \\
    &= \dxi w_0(x)^\ast\left(-\frac{\hbar}{2m}\Partial{^2}{x^2}
      +V(x)\right)w_d(x),
  \end{split}
\end{equation}
where we have used the translational invariance of the problem, the
Hamiltonian can be explicitly rewritten in the Wannier basis,
\begin{equation}
  \label{eq:DiscreteHamilton}
  \mathscr{H}=-\sum_{j,d} J_d(|w_j\rangle\langle w_{j+d}|+|w_{j+d}\rangle\langle w_j|).
\end{equation}
So far no approximations were done except for the omission of the
higher bands, which can be reintroduced in the formalism if
needed. The Hilbert space is now represented by a discrete basis of
localized Wannier states, and it is only a matter of re-interpretation
to read Eq.(\ref{eq:DiscreteHamilton}) as a discrete lattice model of
the problem.

To interpret $J_{d}$ in terms of the band structure, we insert the
expression for the Wannier functions (\ref{eq:WannierFunction}) into
the integral for $J_d$ and obtain
\begin{multline}
  \label{eq:1}
    J_d=-\frac{1}{M}\sum_{k,q}\dxi
    \phi_q(x)^*\left(-\frac{\hbar^2}{2m}\Partial{^2}{x^2}
      +V(x)\right)\\
    \times\phi_k(x)e^{ikad}=-\frac{1}{M}\sum_{k}E(k)e^{ikad}.
\end{multline}
where we observe that the $J_d$ amplitudes are the Fourier
coefficients of the band structure. The band structure is an even
function of $k$ and therefore
\begin{equation}
  \label{eq:2}
  E(k)=-J_0-2\sum_{d>0} J_d\cos(kad).
\end{equation}
Without loss of generality we can set $J_0=0$ in the following.

\begin{figure}
  \centering
 \includegraphics[width=\columnwidth]{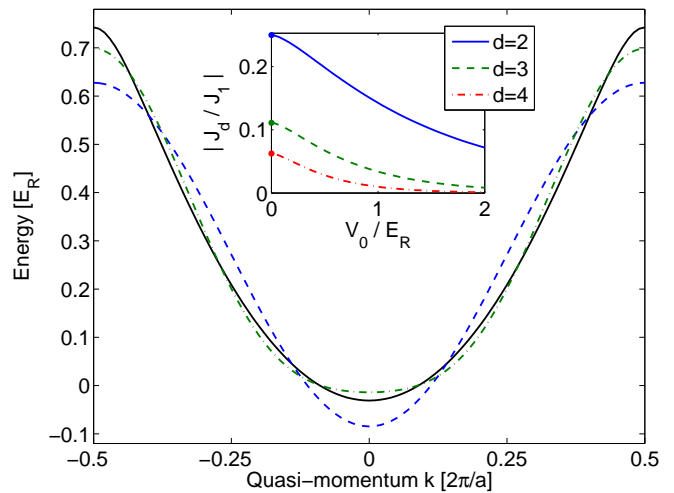}
 \caption{Band structure for a lattice with $V_0=0.5E_R$. Solid line:
   Exact band structure. Dashed line: Apparent band structure in a
   discrete lattice model including only nearest neighbor tunnelling
   (exact only in the large $V_0$ limit). Dash-dotted line: Band
   structure in a discrete lattice model including both $J_1$ and
   $J_2$.  Insert: The importance of higher $J_d$ coefficients as
   function of the lattice strength $V_0$. The filled circles indicate
   the Fourier coefficients of the free space kinetic energy spectrum
   in the first Brillouin zone.}
 \label{fig:BandStructure}
\end{figure}

The application of lattice models which include only nearest neighbor
tunnelling couplings only provide us with a simple cosine shaped band
structure, as depicted in Fig. \ref{fig:BandStructure} (dashed line).
Including more $J_{d}$ terms, the $J_d$-values are compared to $J_1$
in the insert of Fig.  \ref{fig:BandStructure}, gives a more accurate
energy dispersion as indicated by the dash-dotted line in the
figure. As the lattice strength is ramped up in the limit of strong
potentials $V_0>>E_R$, the importance of the higher terms decreases
and only $J_1$ differs significantly from zero.  For $J_2$ to be less
than 1\% of $J_1$, $V_0$ has to exceed $5.32E_R$ \cite{jaksch2005cah}.

We observe that in the opposite limit where $V_0\to 0$, $J_d$
coincides with the Fourier transform of the free space energy
dispersion parabola, i.e. $J_d=-\frac{\hbar^2}{2m}(ad)^{-2}(-1)^d$,
with absolute values plotted as filled circles in the insert of
Fig. \ref{fig:BandStructure}.

Fig. \ref{fig:BandStructure} clearly shows significant and well known
differences between the real band structure and the simple cosine
dependence found for nearest neighbor tunnelling models. These
differences are visible if one tries to perturb and excite motion in
the system, and as we shall see in the following section they also
have important consequences for the states of interacting atoms in the
periodic potential.

\section{Atomic scattering states in the lattice}

We now proceed to the two-body problem and find the eigenstates $\psi(x,y)$ of a pair of particles
moving on the lattice with discrete one-dimensional position coordinates $x$ and $y$, and interacting
by short range interactions. The Hamiltonian of this system is assumed to be of the form
\begin{equation}
  \label{eq:3}
  \mathscr{H}=-\sum_dJ_d(\Delta_x^{(d)}+\Delta_y^{(d)}+4)+U\delta_{x,y}
\end{equation}
where the action of the ``$d^{th}-$Laplacians'' on a wave function are
given by $\Delta_x^{(d)}f(x)=(f(x+ad)+f(x-ad)-2f(x))$ with step size
$ad$. The sum over the discrete Laplacians incorporates the tunnelling
terms \eqref{eq:DiscreteHamilton} which in turn describe both the
kinetic energy and the periodic potential energy of the particles,
i.e. the full dynamics of independent particles in the lattice. The
interaction between particles should be computed as a matrix element
between products of Wannier states of the real physical interaction,
and in case of a short range interaction, represented by a Dirac delta
function $g_{1D}\delta(x-y)$, one would thus find matrix elements in
the product state basis,
\begin{equation}
  \label{eq:UConstants}
  U_{ijkl}=g\od\dxi w_i(x)^*w_j(x)^*w_k(x)w_l(x)
\end{equation}
but we will assume in the following that they fall off rapidly for
Wannier functions situated in different wells, and we apply just the
single on-site Kronecker delta function interaction in (\ref{eq:3}).

We will first discuss the scattering states, where the atoms are asymptotically free. As in normal
scattering theory, the continuum part of the Hamiltonian (\ref{eq:3}), and the sum of the single
particle Hamiltonians of non-interacting particles are unitarily equivalent. This implies that the
energy spectrum is given by simple addition of contributions from two identical band structure
calculations as the ones shown in Fig. 1. We can also solve the scattering problem explicitly and find
the eigenstates, and since we shall need this description to account for the bound states, we shall
proceed and find the two-body wave function in terms of the center-of-mass, $Z=(x+y)/2$, and relative
coordinate, $z=x-y$. Note that the problem of particles moving in the full cosine potential does not
separate in these coordinates, but in the lattice model, the Hamiltonian does not explicitly depend on
the shape of the lattice potential wells (the interaction is built into the ''Laplacian'' and the
Wannier functions making up the discrete states), and separation \emph{is} possible. When we apply the
Hamiltonian \eqref{eq:3} to the wave function Ansatz, $\psi(x,y)=e^{iKZ}\psi_K(z)$, for each value of
$K$ we have to solve the one-body Schr\"odinger equation for the relative position coordinate,
\begin{multline}
  \label{eq:4}
  \left(-2\sum_{d>0}\cos(\tfrac{1}{2}Kda)J_d\left(\Delta_z^{(d)}+2\right)
    +U\delta_{z,0}\right)\psi_K(z)\\=E\psi_K(z).
\end{multline}

\begin{figure}
  \centering
  \includegraphics[width=\columnwidth]{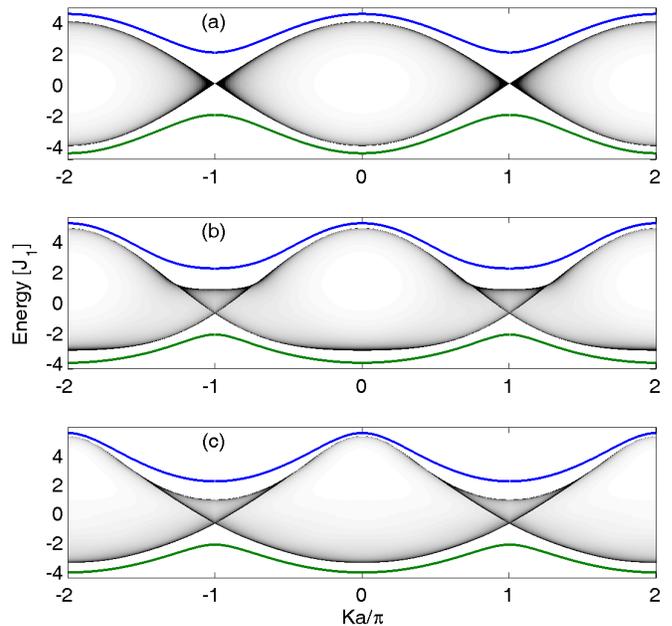}
  \caption{Bound and scattering state spectra for a lattice with
    $V_0=0.5E_R$ and $U=2 J_1$. The shaded bands show the scattering
    states with density of states indicated by the shading. (a)
    Calculation including only $J_1$. (b) Calculation including $J_1$
    and $J_2$. (c) Calculation including $J_d$ up to $d=6$.}
  \label{fig:BoundStates}
\end{figure}

It is readily verified that plane wave functions of the form
$\psi^0_K(z)=e^{ikz}$ are eigenfunctions of the unperturbed
Hamiltonian $H_0=-2\sum_{d>0}\cos(\tfrac{1}{2}Kda)J_d(\Delta_z^{(d)}+2)$
with energies
\begin{equation}
  \label{eq:5}
  \epsilon_{K}(k)=-4\sum_{d>0}J_d\cos(\tfrac{1}{2}Kad)\cos(kad).
\end{equation}
As noted above, the spectrum of scattering states, but not the
scattering wave functions themselves, are independent of the short
range scattering potential, and Eq. \eqref{eq:5} thus provides also
the spectrum of interacting atoms on the lattice.  This spectrum is
displayed in Fig.  \ref{fig:BoundStates}(a)-(c) with inclusion of
different numbers of tunnelling coefficients $J_d$. The different
shades of grey in the figure show the density of states. For fixed
center-of-mass momentum $K$ the density of states has a peak around
the maximal and minimal energy. This is due to the flatness of the
$\cos(kad)$ function in Eq.  \eqref{eq:5} near its extremal values. A
complete degeneracy in $k$ can be observed due to the front factor
$\cos(\tfrac{1}{2}Ka)$ in Eq. \eqref{eq:5}, when only nearest neighbor
coupling is included and $K=\pm\pi/a$. This degeneracy, however, is
lifted by the amount $\Delta\epsilon_{\pm\pi/a}=8\sum_m(-1)^mJ_{2m}$
in Fig. \ref{fig:BoundStates}(b)-(c).

Another pronounced feature in the spectra of
Fig. \ref{fig:BoundStates} is the extra van Hove singularity in the
scattering states for values of the center-of-mass momentum close to
$K=\pm\pi/a$.  To see where it comes from we note that the
non-interaction system separate, so we could have chosen a product
wave function $\Psi=\psi_{q_1}(x)\psi_{q_2}(y)$ with total energy
$E(q_1)+E(q_2)$, where $q_j$ is the quasi-momentum of the individual
atoms and $E$ is the one-body band structure \eqref{eq:2}. If we infer
the center-of-mass quasi-momentum $K=q_1+q_2$ and relative motion
quasi-momentum $k=(q_1-q_2)/2$, we can see that this is indeed the
same energy $\epsilon_{K}(k)$ as derived above
\begin{equation}
  \label{eq:Eq1q2}
  \begin{split}
    \epsilon_K(k)&=E\left(q_1=\tfrac{K}{2}+k\right)+E\left(q_2=\tfrac{K}{2}-k\right)\\
    &=-2\sum_{d>0} J_d\left(\cos\left(\tfrac{K}{2}+k\right)ad +
      \cos\left(\tfrac{K}{2}-k\right)ad\right) \\
    &=-4\sum_{d>0} J_d\cos\tfrac{Kad}{2}\cos kad.
  \end{split}
\end{equation}

\begin{figure}
  \includegraphics[width=\columnwidth]{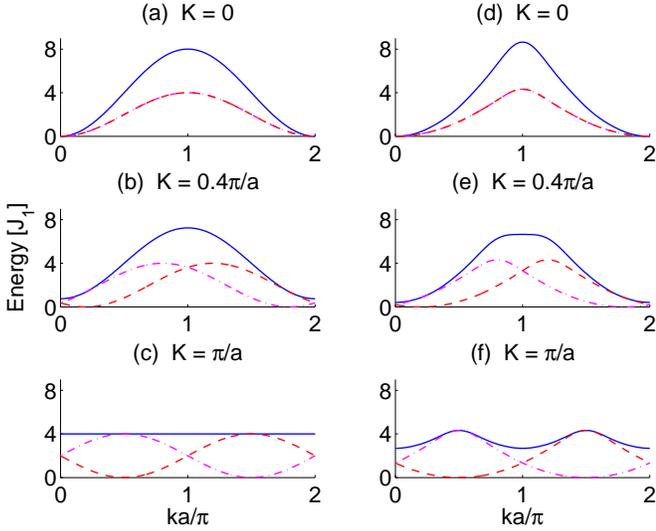}
  \centering
  \caption{Band structure of two-particles $\epsilon_K(k)$ for
    different values of $K$ (solid line) and the band structure of the
    individual atoms, i.e. $E(\tfrac{K}{2}+k)$ and $E(\tfrac{K}{2}-k)$
    (dotted lines). In (a)-(c) the band structure is approximated by a
    cosine and in (d)-(f) higher terms up to $J_6$ is included in the
    band structure, see Eq. \eqref{eq:2}.}
  \label{fig:BS21}
\end{figure}

The extra van Hove singularities can be understood from
Fig. \ref{fig:BS21}, where the band structure for each atom is plotted
together with the total energy as a function of the relative momentum
$k$.  The extra van Hove singularity around $K=\pm\pi/a$ in
Fig. \ref{fig:BoundStates}(b)-(c) is due to the appearance of a local
minimum in the two-body energy spectrum at $k=\pi/a$. This phenomenon
sets in when $\Partial{^2\epsilon_{K}}{k^2}(k=\pi/a)=0$ as shown in
Fig.  \ref{fig:BS21}(e), where $K=0.4\ \pi/a$. We note that
$\Partial{^2\epsilon_{K}(k)}{k^2}= 0$ and
$\Partial{\epsilon_{K}(k)}{k}=0$ cannot be simultaneously fulfilled
when the band structure is a pure cosine function, i.e., when only
nearest neighbor tunnelling is included in the model. When
$K=\pm\pi/a$, as in figure \ref{fig:BS21}(f), the energy minima at
$k=0$ and $k=\pi/a$ are degenerate and therefore we see only two
maxima in the density of state in Fig.  \ref{fig:BoundStates}(c).

In the next section we include the atom-atom interaction, $V(z)=U\delta_{z,0}$, to account for bound
states.

\section{Repulsively and attractively bound states}

The bound states of the system can be obtained from the Green's
function solution of the Schr\"odinger equation (\ref{eq:4}). We first
solve the problem in the absence of the interaction,
\begin{equation}
  \label{eq:6}
  (E-H_0)G_K^0(E,z)=\delta_{z,0}
\end{equation}
to obtain the unperturbed Green's function. The full Green's function in position space can now be
obtained
\begin{eqnarray}
  \label{eq:LS}
  G_K(E,z)&=&G_K^0(E,z)\nonumber\\
  &&+\int \text{d}z'
  G_K^0(E,z-z')V(z')G_K(E,z')\nonumber\\
  &=&G_K^0(E,z)+ G_K^0(E,z)UG_K(E,z=0)\nonumber\\
  &=&G_K^0(E,z)+ \frac{G_K^0(E,z)UG_K^0(E,z=0)}{1-UG_K^0(E,z=0)}\nonumber\\
  &=&\frac{G_K^0(E,z)}{1-UG_K^0(E,z=0)}.
\end{eqnarray}
Whenever there is a bound state $G_K$ has a pole, therefore the bound
state energies are determined solely by the interaction strength $U$
and the unperturbed Green's function
\begin{equation}
  \label{eq:BoundStateEq}
  1=UG_K^0(E_b,z=0).
\end{equation}
The unperturbed Green's function, in turn, is readily obtained from
the momentum space representation, $G_K^0(E,z)=(2\pi)^{-1}\dki
G_K^0(E,k)e^{ikz}$,
\begin{equation}
  \label{eq:7}
  G_K^0(E,k)=\frac{1}{E-\epsilon_{K}(k)+i\eta},\quad (\eta\to 0^+).
\end{equation}
The integral of Eq. \eqref{eq:7} can be found analytically if
$\epsilon_K(k)$ is a simple cosine, and in the general case it can be
computed numerically and the solution to Eq. \eqref{eq:BoundStateEq}
can be found by a simple numerical search.

We can combine Eqs. \eqref{eq:6} and \eqref{eq:BoundStateEq} to obtain
the relation
\begin{equation}
  \label{eq:9}
  \begin{split}
    (E-H_0)G^0_K(E_b,z)&=\frac{1}{G_K^0(E_b,z=0)}\delta_{z,0}G_K^0(E_b,z)\\
    &=U\delta_{z,0}G_K^0(E_b,z),
  \end{split}
\end{equation}
This shows that $G_K^0(E_b,z)$ is itself a solution to the
Schr\"odinger equation, i.e., it provides directly the bound state
wave function $\psi_K(z)$, and $G_K^0(E_b,k)$ provides the wave
function $\psi_K(k)$ in momentum space.

There are discrete bound states for both positive and negative $U$
corresponding to repulsive and attractive interaction,
respectively. The energy of the repulsively and attractively bound
atom pairs lie above and below the scattering continuum as plotted in
Fig. \ref{fig:BoundStates}. The repulsively bound pair state was
recently experimentally demonstrated by Winkler et al.
\cite{winkler2006rba}.

If we only include $J_1$ we have the following symmetry in the
continuum spectrum, Eq. \eqref{eq:5},
\begin{eqnarray}
  \label{eq:symcont}
  &\epsilon_{K}(k)=-\epsilon_{K}(k+\pi/a).
\end{eqnarray}
This implies, according to Eq. \eqref{eq:7}, when $E$ is outside the
scattering continuum,
\begin{eqnarray}
  \label{eq:sym2G}
  &G_K^0(E,k)=-G_{K}^0(-E,k+\pi/a),
\end{eqnarray}
and hence
\begin{equation}
  \begin{split}
    1=&UG_K^0(E_{ab}(K),z=0)\\
    =&(-U)G_{K}^0(-E_{ab}(K),z=0)\\
    =&(-U)G_{K}^0(E_{rb}(K),z=0),
  \end{split}
\end{equation}
which relates the repulsively ($rb$) and attractively ($ab$) bound
atom pair energies:
\begin{eqnarray}
  \label{eq:8}
  &E_{ab}(K)=-E_{rb}(K).
\end{eqnarray}
Eq. \eqref{eq:9} now implies, if we include only $J_1$, that the
attractively and repulsively bound wave functions are related by
\begin{eqnarray}
  \label{eq:daVeigaSymmetry}
  &\psi_K^{ab}(k)=\psi_K^{rb}(k+\pi/a).
\end{eqnarray}

This symmetry is also pointed out in \cite{PhysRevE.66.016130}, but it is
important to note that when the higher $J_d$ tunnelling couplings are
included, Eq.  \eqref{eq:symcont} and hence the following equations
(\ref{eq:8},~\ref{eq:daVeigaSymmetry}) are no longer valid.

\begin{figure}
  \centering
  \includegraphics[width=\columnwidth]{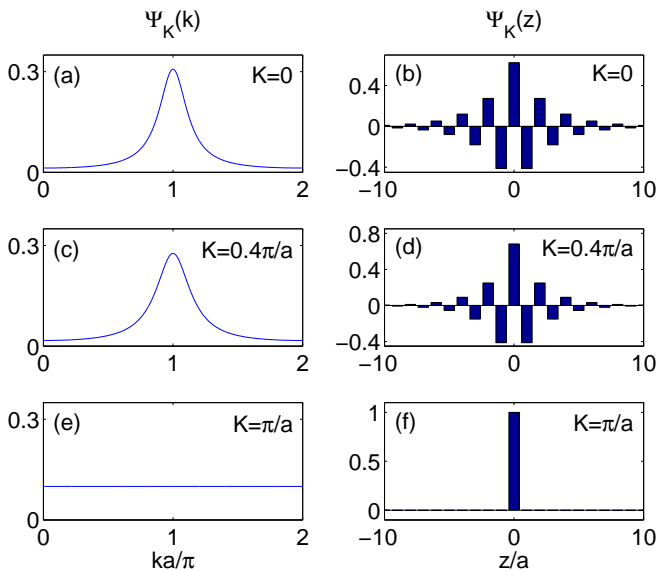}
  \caption{Repulsively bound states obtained with nearest neighbor
    interaction only. First column shows $\psi_K(k)$ and the second
    $\psi_K(z)$ both obtained from Eqs. (\ref{eq:7},~\ref{eq:9}), see
    text for details.}
  \label{fig:RBNearest}
\end{figure}

\begin{figure}
  \centering
  \includegraphics[width=\columnwidth]{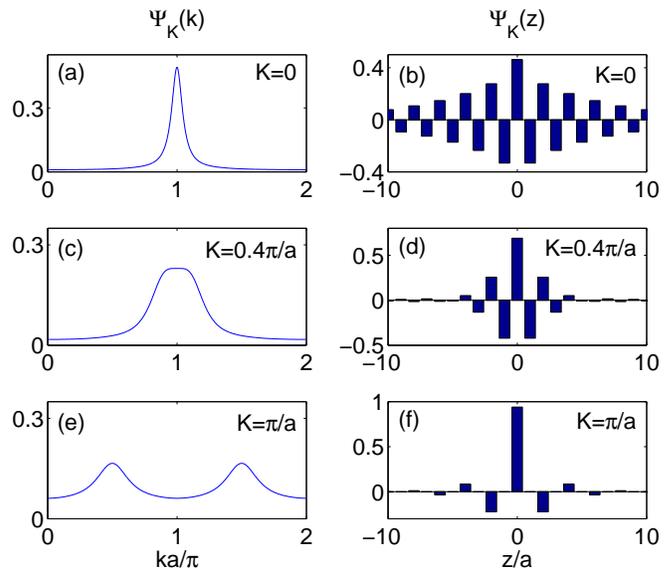}
  \caption{Repulsively bound states obtained with six nearest neighbor
    interactions included. First column shows $\psi_K(k)$ and the
    second $\psi_K(z)$ both obtained from
    Eqs. (\ref{eq:7},~\ref{eq:9}), see text for details.}
  \label{fig:RBNearest2}
\end{figure}

The shape of the band structure is altered when $J_{d>1}$ are
included, but more importantly the wave functions are altered
significantly, compare Figs. \ref{fig:RBNearest} and
\ref{fig:RBNearest2}.

In Fig. \ref{fig:RBNearest} only $J_{d=1}$ is included. As one would
expect the bound states are composed of the quasi-momentum components
lying closest in energy to the bound state. In the region around $K=0$
this correspond to relative momentum $k=\pi/a$ and the formation of
bound pairs can in this region be explained by the fact that the peak
in $k$-distribution corresponds to the individual atoms having
quasi-momentum $q_1=q_2=\pi/a$. Here the effective mass is negative
and therefore a repulsive force causes attraction
\cite{2006JPhA...39L.667M}. As $K$ approaches $\pm\pi/a$ the
scattering states with different relative momenta become degenerate in
energy and the momentum distribution of the bound state goes from a
peaked function Fig. \ref{fig:RBNearest}(a) to a constant function
\ref{fig:RBNearest}(e), which corresponds to a delta function in
relative coordinate, \ref{fig:RBNearest}(f).

In Fig. \ref{fig:RBNearest2}, $J_d$ up to $d\leq 6$ has been included
in obtaining the $k$- and $r$-distributions of the repulsively bound
states.  As discussed previously; when the extra van Hove singularity
sets in Fig. \ref{fig:BS21}(e), the two-body spectrum broadens around
$k=\pi/a$ and this can also be observed in the decomposition of the
bound states fig. \ref{fig:RBNearest2}(c).  In
Fig. \ref{fig:RBNearest2}(e) we clearly see the consequences of
lifting the $k$-degeneracy at $K=\pm\pi/a$. The momentum distribution
is no longer constant but instead peaked around the two momenta
$k=\pm\pi/2a$ which compose the top of the scattering band, as can be
seen in Fig. \ref{fig:BS21}(f).

In the experiment by Winkler et al. \cite{winkler2006rba} the atoms
were prepared in the $K=0$ bound state. We suggest that by exposing
the system to a force field the bound pairs can be prepared in any
$K$-state. This can be realised in practice by chirping the lattice
lasers so the lattice potential is accelerated in the laboratory frame
\cite{Dahan96}. To investigate the effect on the two-atom states, we
make the new time dependent Ansatz
\begin{equation}
  \label{eq:BOAnsatz}
  {\Psi}(x,y,t)=e^{iK(t)\frac{x+y}{2}}{\psi}(x-y,t)
\end{equation}
with a time dependent center-of-mass quasi-momentum, $K(t)$. Inserting
this Ansatz into the time dependent Schr\"odinger equation with the
Hamiltonian \eqref{eq:3} and the linear term due to the force,
$-F\cdot (x+y)$, we obtain
\begin{multline}
  \label{eq:OBligning}
  i\hbar\Partial{}{t}\psi(z,t)=\\\bigg(
  -2\sum_{d>0}\cos(\tfrac{1}{2}K(t)da)J_d\left(\Delta_z^{(d)}+2\right)
  +U\delta_{z,0}\bigg)\psi(z,t)
\end{multline}
if we choose
\begin{equation}
  \label{eq:OBK}
  \dot{K}(t)=\frac{2F}{\hbar}.
\end{equation}
Since there is an energy gap between the bound state and the
scattering states, we may apply the adiabatic approximation to
Eq. \eqref{eq:OBligning} and simply replace $\psi(x-y,t)$ by the
eigenstate $\psi_K(x-y)$ of Eq. \eqref{eq:4} with $K=K(t)$. Provided
we accelerate the lattice slowly enough we can thus explore all the
wave functions $\psi_{K}(x-y)$ of the system with $K=2Ft/\hbar$.

\section{Discussion}
\label{sec:discussion}

The states and energies of particles in periodic potentials depend on
the details of these potentials.  We have in this paper shown that the
lowest band in the single particle band structure in such a potential
can be described very accurately with a discrete lattice model with
tunnelling couplings to a sufficiently high number of remote
sites. Such modelling may be particularly important when one studies
scattering processes where energy and momentum are conserved. We have
previously studied the break-up of a single condensate as a four wave
mixing process \cite{PRA:041602}, which cannot take place if the band
structure is a simple cosine function. In the present work we
described the continuum of scattering states of atoms interacting by a
short range attractive or repulsive interaction, and we identified the
two-atom repulsively and attractively bound states.

\begin{figure}
  \centering
  \includegraphics[width=0.8\columnwidth]{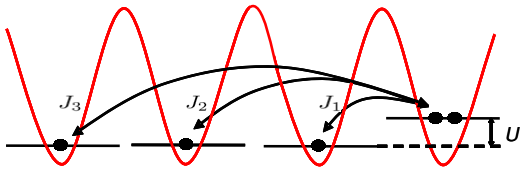}
  \caption{Illustration of the tunnelling and interaction coefficients
    in the Bose-Hubbard Hamiltonian with ``long distance'' tunneling.}
  \label{fig:Picture}
\end{figure}
Simple models for the periodic potentials may be of particular
interest when systems with a large number of particles are studied. In
this case, one has recourse to second quantization, and the
restriction to a discrete set of Wannier states is a first step
towards an accurate treatment of the many-body properties, e.g., how
the atoms distribute themselves among potential wells. See
ref. \cite{denschlag2006eap} for a whole list of projects for further
studies. So far many problems were dealt with by the Bose-Hubbard
Hamiltonian, \cite{PRL:81-3108, PRA:053601, 1464-4266-5-2-352,
  PhysRevLett.91.080403, jaksch2005cah, 2006cond.mat.10198P}, either
solved exactly numerically or exposed to further simplifying
approximations. Based on the experience of the present work, we
suggest that in cases where collisions with momentum and energy
conservation play a role, one should extend the conventional
Bose-Hubbard Hamiltonian with only nearest neighbor tunnelling to
include also tunnelling terms over longer distances,
$\mathscr{H}=-\sum_{j,d>0}
J_d(\hat{a}_j^\dagger\hat{a}_{j+d}+\hat{a}_{j+d}^\dagger\hat{a}_j)+
U\hat{a}_j^\dagger\hat{a}_j^\dagger\hat{a}_j\hat{a}_j$, as depicted in
Fig. \ref{fig:Picture}. This modified Hamiltonian, or further extended
with non-local interaction terms \cite{mazzarella:013625,
  heiselberg:013628}, would seem to be a good starting point for
detailed studies of collisions and instabilities in moving condensates
in periodic potentials.  We are currently pursuing such studies,
including also the interplay of repulsive binding and real molecular
binding of atoms in optical lattices.


\end{document}